\documentclass[a4paper,12pt]{article}
\usepackage[utf8]{inputenc}
\usepackage{cancel}
\usepackage{ulem}
\usepackage{amsfonts}
\usepackage{amssymb}
\usepackage{graphicx}
\usepackage{amsmath}
\usepackage{enumerate}
\usepackage{mathtools}
\usepackage{subfig}
\usepackage{color}
\usepackage{tikz}\usetikzlibrary{calc}
\usepackage{float}
\usepackage{here}
\usepackage{cite}
\usepackage{mathrsfs}
\usepackage{float,epsfig}
\usepackage{dcolumn}

\usepackage{graphicx}
\usepackage{bm}
\usepackage{amsmath,amssymb,amsthm}
\usepackage[colorlinks=true,linkcolor=blue,citecolor=red]{hyperref}
\newcommand{\be}{\begin{equation}}
\newcommand{\ee}{\end{equation}}
\newcommand{\bea}{\setlength\arraycolsep{2pt} \begin{eqnarray}}
\newcommand{\eea}{\end{eqnarray}}

\def\0{{\sst{(0)}}}
\def\1{{\sst{(1)}}}
\def\2{{\sst{(2)}}}
\def\3{{\sst{(3)}}}
\def\4{{\sst{(4)}}}
\def\5{{\sst{(5)}}}
\def\6{{\sst{(6)}}}
\def\7{{\sst{(7)}}}
\def\8{{\sst{(8)}}}
\def\sst#1{{\scriptscriptstyle #1}}

\makeatletter \@addtoreset{equation}{section}

\usepackage{multirow}
\definecolor{lime}{HTML}{A6CE39}
\newcommand{\orcidicon}{%
    \begin{tikzpicture}
    \draw[lime, fill=lime] (0,0)
        circle [radius=0.16]
        node[white] {{\fontfamily{qag}\selectfont \tiny ID}};
    \draw[white, fill=white] (-0.0625,0.095)
        circle [radius=0.007];
    \end{tikzpicture}   \hspace{-2mm}
}
\newcommand\orcidAdil{{\href{https://orcid.org/0000-0001-7623-5541}{\orcidicon}}}
\newcommand\orcidHajar{{\href{https://orcid.org/0000-0001-9510-4248}{\orcidicon}}}              
\newcommand\orcidHasan{{\href{https://orcid.org/0000-0001-7408-0910}{\orcidicon}}}
\newcommand\orcidMohamed{{\href{https://orcid.org/0000-0003-1185-0062}{\orcidicon}}}

\begin{document}
%

\title{\normalsize
{\bf \Large	  Light  Deflection by Rotating Regular   Black Holes    with a Cosmological  Constant   }}
\author{ \small   A. Belhaj\orcidAdil\!\! $^{1}$\footnote{a-belhaj@um5r.ac.ma},  H. Belmahi\orcidHajar\!\!$^{1}$\footnote{hajar\_belmahi@um5.ac.ma},  M. Benali\orcidMohamed\!\!$^{1}$\footnote{mohamed\_benali4@um5.ac.ma}, H. El Moumni\orcidHasan\!\!$^{2}$\thanks{h.elmoumni@uiz.ac.ma}
	\footnote{ Authors in alphabetical order.}
	\hspace*{-8pt} \\
	{\small $^1$ D\'{e}partement de Physique, Equipe des Sciences de la mati\`ere et du rayonnement, ESMaR}\\
{\small   Facult\'e des Sciences, Universit\'e Mohammed V de Rabat, Rabat,  Morocco} \\
	{\small $^{2}$  EPTHE, D\'{e}partement de Physique, Facult\'e des Sciences,   Universit\'e Ibn Zohr, Agadir, Morocco} 
} \maketitle

 \maketitle
\begin{abstract}
Using the  Gauss-Bonnet theorem,   we compute  and examine  the  deflection angle of light rays by rotating  regular black  holes with a cosmological  constant.  By the help of   optical geometries, we first deal with the Hayward black holes with  cosmological  contributions.   Then, we reconsider the study  of  the Bardeen solutions.  We  inspect  the cosmological constant effect  on the deflection angle of light rays.    Concretely,  we find extra  cosmological correction terms generalizing certain  obtained  findings.   Using  graphical analysis, we provide a comparative  discussion  with respect to the Kerr solutions.  The results confirm that the non-linear electrodynamic charges affect the space-time geometry by decreasing  the deflection angle of light rays  by such cosmological   black holes.
		{\noindent}
	
	\end{abstract}

\newpage
\section{Introduction}
Black hole physics has received a remarkable interest  due  to  recent detections and observational  findings supporting the obtained  theoretical investigations.  Precisely,  Even Horizon  Telescope (EHT) has   provided an image of  a black hole \cite{BW9,A2, A3,A4}. This discovery  has  been followed by  significant  efforts   reserved  to the investigation  of  various  black hole  aspects \cite{hhh,hh}. These involve the thermodynamics and the optical behaviors  encoding  many interesting  data of the associated physics.   In connections with Anti de Sitter geometries,   the thermodynamic quantities have been approached  by linking the cosmological  constant with the pressure \cite{12,F,F1, E}.  Based on such an interplay,  certain phase transitions, including the Hawking-Page one,  have been dealt with extensively in many places including in \cite{D}. It has been observed many similarities with fluid physical systems \cite{f2,E1}.  Besides that, optical properties of the black holes have been also   studied  by focusing  on  two relevant  notions, being the shadow and the  deflection angle of light rays. The  first concept  has been  discussed in terms of one dimensional real curves \cite{BW12,H,B12,BC}. In this way,  the corresponding size and shape  have been controlled by the involved black hole parameters \cite{I,Xa,J,RC,K}.   The second notion concerns the  deflection angle of light rays  by black holes which   has been investigated using different methods\cite{BW13,BW15,BW17,TR1,TR3}. A priori,  there are two famous ones.  One is relied on an elliptic formalism involving non-trivial  elliptic functions including the hypergeometric  ones, explored to compute the so-called total deflection angle\cite{M5,N5}. The second method, which will be exploited here,  is based on the  Gauss-Bonnet theorem results \cite{BW18,BW23,TR2}.
Precisely,  Gibbons and Werner  suggested a nice  road   to compute the deflection angle of light rays  by black holes \cite{BW16}.  In this regard,  many models have been discussed including regular black hole  geometries \cite{JK}.  These solutions have been  interpreted  in terms of     gravitational    backgrounds    with non-linear electromagnetic charges. 
The key interest on these  black holes comes from  the understanding of  certain singularities  thanks to the  pioneer works of Penrose and Hawking \cite{65,66}.   Then,  Bardeen discovered a regular black hole obeying the weak energy condition.  After that, many Bardeen-like black hole solutions  have been investigated   where the  irregularity  has been  linked to topological changes. This has been exploited to  provide a  possibility to  build  spaces with a maximum curvature inside the black holes \cite{67,68}.  Moreover, other regular black hole models have been  suggested  by coupling  the Einstein’s theory with nonlinear electrodynamics \cite{71}.  It has been shown that the  regular black hole interior solutions have been  found in  many gravity models  like Loop Quantum Gravity \cite{72}.  
The   black hole  solutions  of  Bardeen
and   Hayward  have been extensively  studied by examining either thermodynamics properties or shadow optical ones \cite{Alt, Mol}. 

The aim of this work is to contribute to such activities  by investigating the deflection angle of light rays  by regular black holes with a cosmological constant.   Using the techniques explored in \cite{BW18,BW23,TR2}, we compute and examine  such an optical quantity of the   rotating Hayward  regular black holes and Bardeen regular black holes in  terms of existing  parameters.  Taking  the  finite distance  contributions,     we   inspect the effect  of such parameters. A particular emphasis puts   on     the cosmological constant influence.  Then, we provide a comparison study with respect  to the Kerr solutions with negative and positive cosmological constant values.  This   finding confirms that  the non-linear electrodynamic charges affect the space-time geometry by decreasing  the  deflection angle of light rays  near such  black holes with a cosmological constant.

This paper is organized  as follows.  In section 2, we present the formalism needed to   compute  the deflection angle of light rays.  In section 2, we  consider the rotating regular Hayward black hole  with a cosmological  constant.   In section 4, we deal with   the rotating regular Bardeen black holes with cosmological contributions.  In section 5, we provide a comparative study  by elaborating graphical analysis using results of the Kerr solutions. The last  section concerns concluding remarks.

\section{Deflection angle  formalism}
In this section, we give  a concise review on the computations of the light  deflection angle around a black hole.   An examination shows  that various  methods have been  proposed and exploited for many solutions.  The  known one has been based on  the Gauss-Bonnet theorem results  being extensively used to approach several   black hole backgrounds \cite{BW18}. In the present work, we  will  exploit these  findings to  examine   such an optical quantity where   the observer and the source   are placed at finite distances.   In particular,  the calculations relay on the method developed in \cite{BW23}.  In this way, the deflection angle of light rays  can be  expressed  via the relation
 \begin{equation}
 \Theta=\Psi_{R}-\Psi_{S}+\phi_{SR}
 \label{a1}
 \end{equation}
where  $\Psi_{R}$ and $\Psi_{S}$    are the  angles between the light rays and the radial direction at the observer and  the source position,  respectively.  It is denoted that    $\phi_{SR}$   is the longitude separation angle, which will be specified  later on\cite{BW18}.  To write down the associated  formula, an  unit tangential vector along the light  will be used.  This vector denoted $e^i$  is linked to the above angle as follows
\begin{equation} 
\label{L2}
(e^r,e^\theta,e^\phi)=\epsilon\left(\frac{dr}{d\phi},0,1\right)
\end{equation}
where  $\epsilon$     is a radial quantity which  can be determined from the    metric black  hole solutions. Putting such a metric as follows 
\begin{equation}
ds^2=-A(r,\theta)dt^2+B(r,\theta)dr^2+C(r,\theta)d\theta^2+D(r,\theta)d\phi^2-2H(r,\theta)dt d\phi,
\end{equation}
it is possible to get  its expression.  Concretely, it is given by 
\begin{equation}
\label{L3}
\epsilon=\frac{A(r)D(r)+H^2(r)}{A(r)(H(r)+A(r)b)}
\end{equation}
where $b$ is the impact parameter  considered  as a ratio of the two constants of motion  derived  from the orbit equation. A simplified form can be determined    by considering the equatorial plane.   Indeed,  these  two conserved quantities   read as 
\begin{equation}
E= A(r)\dot  t +H(r)\dot \phi \qquad    L= D(r)\dot  \phi-H(r)\dot t
\label{labe}
\end{equation}
where the derivative with respect to the affine parameter has been used.   They represent  the energy  and  the angular momentum of the test  particle, respectively.  Considering a constant $t$  of the space-time metric,    a  2-dimensional curved space   can be built  being relevant  in the calculations of the  light deflection angle.  In the equatorial plane,  its  element line should take the following form
\begin{equation}
dl^2\equiv \gamma_{ij}dx^{i}dx^{j}
\label{labe}
\end{equation}
where $\gamma_{ij}$ is  a spatial metric encoding the involved  physical parameters. In this space, the     deflection  angle of light rays  can be  measured from the radial direction.    The above  angle  can be determined by   taking   the particular angle $\theta=\frac{\pi}{2}$ and using the relation  
\begin{equation}
\label{L1a}
\cos\Psi\equiv \gamma_{ij}e^iR^j,
\end{equation}
where   $R^j$ are  the components of  a radial vector given by $(\frac{1}{\sqrt{\gamma_{rr}}},0,0)$.  Exploiting  Eq.(\ref{L2}) and Eq.(\ref{L1a}),     the $\sin\Psi$  term  can be obtained  from  the following relation
\begin{equation}
\sin\Psi=\frac{H(r)+A(r)b}{\sqrt{A(r)D(r)+H^2(r)}}.
\label{EE}
\end{equation}
With this formalism in hand,  we  can  calculate the deflection angle of light rays by rotating  regular black holes in cosmological backgrounds by considering the retrograde situation. This will be the aim of the forthcoming sections.   In particular, we consider two models followed by a comparative discussion.
\section{Deflection angle of light rays by  cosmological Hayward black hole solutions }
In this section,    we   investigate the deflection angle of light rays around the rotating  Hayward black holes with a cosmological constant.  As mentioned before,  the calculations    are based on the metric form. In  the   Boyer-Lindquist coordinates,   this is given  via the following  line element
\begin{equation}
ds^{2}=-\frac{\Delta_{r}}{\Sigma}\left(dt-\frac{a \sin^{2}\theta}{\Xi}d\phi\right)^{2}+\frac{\Sigma}{\Delta_{r}}dr^{2}+\frac{\Sigma}{\Delta_{\theta}}d\theta^{2}+\frac{\Delta_{\theta} \sin^{2}\theta}{\Sigma}\left(a dt - \frac{r^{2}+a^{2}}{\Xi}d\phi\right)^{2}.
\end{equation}
The physical terms appearing in  such a metric    can be  expressed as follows  
\begin{eqnarray}
\Delta_{r}&=& \left(r^{2}+a^{2}\right)\left(1-\frac{r^{2}\Lambda}{3}\right)-2M\left(\frac{r^{3}}{r^{3}+g_{h}^{3}}\right)r, \hspace{1cm} \Delta_{\theta}= 1+\frac{a^{2}\Lambda}{3}\cos^{2}\theta, \\ [6px]
\Xi &=& 1+\frac{a^{2}\Lambda}{3}, \hspace{1cm}  \Sigma = r^{2}+a^{2} \cos^{2}\theta.
\end{eqnarray}
 Here,   $M$ and $a$  represent the mass and the rotating parameter, respectively.  $g_h$   indicates   the charge of the 
non-linear electrodynamics and $\Lambda$  is   the cosmological constant.   Using the impact parameter given by $ b=\frac{L}{E}$ and changing the variable $r$ to $\frac{1}{u}$, we can get 
\begin{eqnarray}
\left( \frac{du}{d\phi}\right)^{2} &=&\frac{1}{b^2}-\frac{4 a M u}{b^3}-u^2+2 M u^3+\frac{\Lambda }{3}-\frac{2 a \Lambda }{3 b^3 u^2}-\frac{8 a \Lambda  M}{3 b^3 u}+\frac{4 a M u^4 g_h^3}{b^3}-2 M u^6 g_h^3\notag\\ &+&\frac{8 a \Lambda  M u^2 g_h^3}{3 b^3}+\mathcal{O}({M^2,\Lambda^2,a^2, g_{h}^4}),
\end{eqnarray}
where one has considered  only the retrograde  solution.  In the beginning,  one should calculate the separation angle integral.    Putting $ \left( \frac{du}{d\phi}\right)^{2} =F(u)$,  this can be determined  via the following   computation 
\begin{equation}
\phi_{RS} = \int^R_S d\phi= \int^{u_0}_{u_S}\frac{1}{\sqrt{F(u)}}du +\int^{u_0}_{u_R}\frac{1}{\sqrt{F(u)}}du , 
\end{equation}
where $ u_S $ and $  u_R$ are the inverse of the source and the observer distance from the black hole and where $  u_0$ is the inverse of  the closest approach $r_{0}$. Considering the  
weak field and the slow rotation approximations, the impact parameter can be related to $u_{0}$ as follows
\begin{equation}
b = \frac{1}{u_{0}}+M-2aM u_{0}+\mathcal{O}({M^2,\Lambda,a^2, g_{h}}).
\end{equation}

Performing appropriate calculations, we obtain 
\begin{eqnarray}
\phi_{RS} &=& \phi_{RS}^{Kerr}+\left(\frac{u_R \left(2 b^4 u_R^4+5 b^2 u_R^2-15\right)}{\sqrt{1-b^2 u_R^2}}+\frac{u_S \left(2 b^4 u_S^4+5 b^2 u_S^2-15\right)}{\sqrt{1-b^2 u_S^2}} \right) \frac{M g_h^3}{8 b^3}\notag \\ &+& \left(\frac{u_R \left(b^2 u_R^2-3\right)}{\sqrt{1-b^2 u_R^2}}+\frac{u_S \left(b^2 u_S^2-3\right)}{\sqrt{1-b^2 u_S^2}} \right) \frac{a M g_h^3}{b^4}\notag \\ &-&\left(\frac{u_R \left(3 b^4 u_R^4-20 b^2 u_R^2+15\right)}{\left(1-b^2 u_R^2\right){}^{3/2}}+\frac{u_S \left(3 b^4 u_S^4-20 b^2 u_S^2+15\right)}{\left(1-b^2 u_S^2\right){}^{3/2}} \right) \frac{\Lambda  M g_h^3}{12 b}\notag \\ &+& 
\left( \pi-\arcsin\left(b u_R\right)-\arcsin\left(b u_S\right)\right) \left(-\frac{15 M g_h^3}{8 b^4}+ \frac{3 a M g_h^3}{b^5}-\frac{5 \Lambda  M g_h^3}{4 b^2}\right) \notag \\ &+& \left(\frac{ u_R}{ \sqrt{1-b^2 u_R^2}}+\frac{ u_S}{ \sqrt{1-b^2 u_S^2}}\right)\frac{b^3 \Lambda}{6} +\left(\frac{b \left(2-3 b^2 u_R^2\right)}{2 \left(1-b^2 u_R^2\right){}^{3/2}}+\frac{b \left(2-3 b^2 u_S^2\right)}{2 \left(1-b^2 u_S^2\right){}^{3/2}}\right)\frac{M\Lambda}{3}\notag \\ &+& \left( \frac{1-2 b^2 u_R^2}{u_R \sqrt{1-b^2 u_R^2}}+\frac{1-2 b^2 u_S^2}{u_S \sqrt{1-b^2 u_S^2}}\right) \frac{a\Lambda}{3}+ \mathcal{O}({M^2,\Lambda^2,a^2, g_{h}^4},M a\Lambda,M a\Lambda  g_{h}^3 ),
\end{eqnarray}
where the Kerr term is given by
\begin{eqnarray}
 \phi_{RS}^{Kerr} &=& \phi_{RS}^{SW} - \left(\frac{1}{\sqrt{1-b^2 u_R^2}}+\frac{1}{ \sqrt{1-b^2 u_S^2}}\right)\frac{2aM}{b^2}.
\end{eqnarray}
In this equation,   the Schwarzschild   term  is \begin{eqnarray}\phi_{RS}^{SW}=\pi -\arcsin\left(b u_R\right)-\arcsin\left(b u_S\right)+ \left(\frac{2-b^2 u_R^2}{ \sqrt{1-b^2 u_R^2}}+\frac{2-b^2 u_S^2}{\sqrt{1-b^2 u_S^2}}\right) \frac{M}{b}.\end{eqnarray}
To get the remaining terms appearing in the light deflection angle expression,   one should   identify the $\Psi$ terms. By using the Eq.(\ref{EE}), we find 
\begin{eqnarray}
\sin \Psi&=&b u-b M u^2+2 a M u^2- \left(\frac{b M}{6} - \frac{a  }{3 u}+\frac{b }{6 u}\right)\Lambda +\left( b  u^5  -2 a u^5 +\frac{b \Lambda   u^3 }{6}\right)M g_h^3 \notag \\ &+&\mathcal{O}({M^2,\Lambda^2,a^2, g_{h}^4},M a\Lambda,M a\Lambda  g_{h}^3 ).
\end{eqnarray}
This relation  produces 
\begin{eqnarray}
\Psi_{R}&-&\Psi_{S}=\Psi_{R}^{Kerr}-\Psi_{S}^{Kerr}+\left( \frac{b u_R^5}{\sqrt{1-b^2 u_R^2}}+\frac{b u_S^5}{\sqrt{1-b^2 u_S^2}}\right) M g_h^3\notag \\ &-&\left( \frac{u_R^5}{\sqrt{1-b^2 u_R^2}}+\frac{u_S^5}{\sqrt{1-b^2 u_S^2}}\right) 2 a M g_h^3-\left( \frac{u_R^3 \left(2 b^2 u_R^2-1\right)}{\left(1-b^2 u_R^2\right){}^{3/2}}+\frac{u_S^3 \left(2 b^2 u_S^2-1\right)}{\left(1-b^2 u_S^2\right){}^{3/2}}\right) \frac{ b \Lambda  M g_h^3}{6}\notag \\ &-&\left( \frac{1}{u_R \sqrt{1-b^2 u_R^2}}+\frac{1}{u_S \sqrt{1-b^2 u_S^2}}\right) \frac{b \Lambda}{6}-\left( \frac{2 b^2 u_R^2-1}{\left(1-b^2 u_R^2\right){}^{3/2}}+\frac{2 b^2 u_S^2-1}{\left(1-b^2 u_S^2\right){}^{3/2}}\right) \frac{b \Lambda M }{6}\notag \\ &+&\left( \frac{1}{u_R \sqrt{1-b^2 u_R^2}}+\frac{1}{u_S \sqrt{1-b^2 u_S^2}}\right) \frac{a \Lambda}{3}+  \mathcal{O}({M^2,\Lambda^2,a^2, g_{h}^4},M a\Lambda,M a\Lambda  g_{h}^3 ),
\end{eqnarray}
where one has 
\begin{eqnarray}
\Psi_{R}^{Kerr}&-&\Psi_{S}^{Kerr}=\Psi_{R}^{SW}-\Psi_{S}^{SW}+\left( \frac{u_R^2}{\sqrt{1-b^2 u_R^2}}+\frac{u_S^2}{\sqrt{1-b^2 u_S^2}}\right) 2 a M,
\end{eqnarray}
and where  one has found 
\begin{eqnarray}
\Psi_{R}^{SW}-\Psi_{S}^{SW}=\left( \arcsin\left(b u_R\right)+\arcsin\left(b u_S\right)-\pi\right) -\left( \frac{u_R^2}{\sqrt{1-b^2 u_R^2}}+\frac{u_S^2}{\sqrt{1-b^2 u_S^2}}\right)M b .
\end{eqnarray}
Combining the above equations, we get  an expression of  the  light deflection angle given by 

\begin{eqnarray}
\alpha_{h}&=& \left(\sqrt{1-b^2 u_R^2}+\sqrt{1-b^2 u_S^2} \right)\frac{2 M}{b}- \left( \sqrt{1-b^2 u_R^2}-\sqrt{1-b^2 u_S^2} \right)\frac{2 a M}{b^2}\notag \\ &-& \left( \frac{1-b^2 u_R^2}{u_R \sqrt{1-b^2 u_R^2}}+\frac{1-b^2 u_S^2}{u_S \sqrt{1-b^2 u_S^2}}\right) \frac{\text{b$\Lambda $}}{6}+\left( \frac{1}{\sqrt{1-b^2 u_R^2}}+\frac{1}{\sqrt{1-b^2 u_S^2}}\right)\frac{b \Lambda  M}{6}\notag \\ &+& \left(\frac{\sqrt{1-b^2 u_R^2}}{u_R}+\frac{\sqrt{1-b^2 u_S^2}}{u_S}\right) \frac{2 a \Lambda }{3}+\left( \frac{u_R \left(2 b^4 u_R^4+b^2 u_R^2-3\right)}{\sqrt{1-b^2 u_R^2}}+\frac{u_S \left(2 b^4 u_S^4+b^2 u_S^2-3\right)}{\sqrt{1-b^2 u_S^2}}\right) \frac{5 M g_h^3}{8 b^3}\notag \\ &-& \left(\frac{u_R \left(2 b^4 u_R^4+b^2 u_R^2-3\right)}{b^4 \sqrt{1-b^2 u_R^2}}+\frac{u_S \left(2 b^4 u_S^4+b^2 u_S^2-3\right)}{b^4 \sqrt{1-b^2 u_S^2}}\right) \frac{a M g_h^3}{b^4} \notag \\ &+&\left(\frac{u_R \left(7 b^2 u_R^2-15\right)}{\sqrt{1-b^2 u_R^2}}+\frac{u_S \left(7 b^2 u_S^2-15\right)}{\sqrt{1-b^2 u_S^2}} \right)\frac{\Lambda  M g_h^3}{12 b}\notag \\ &+&\left(\pi -\arcsin\left(b u_R\right)-\arcsin \left(b u_S\right)\right)\left(-\frac{15 M g_h^3}{8 b^4}+ \frac{3 a M g_h^3}{b^5}-\frac{5 \Lambda  M g_h^3}{4 b^2}\right) \notag \\ &+&\mathcal{O}({M^2,\Lambda^2,a^2, g_{h}^4},M a\Lambda,M a\Lambda  g_{h}^3 )   .
\end{eqnarray}
This form   can be reduced to a simplified  one  using certain  convenable  approximations.  Taking $u_Sb<<1$ and $u_Rb<<1$,     we can get an expression involving divergent terms coupled to  the cosmological   contributions. These terms  should be  existed  to show the  cosmological background dependance. The desired deflection angle of light rays  is found to be 
\begin{eqnarray}
\alpha_{h} &=& \frac{4 M}{b}-\frac{4 a M}{b^2}-\frac{15 \pi  M g_h^3}{8 b^4}+\frac{3 \pi  a M g_h^3}{b^5}-\frac{5 \Lambda  \text{$\pi $M} g_h^3}{4 b^2} +\frac{b \Lambda  M}{3}-\left(\frac{1}{u_R}+\frac{1}{u_S}\right)\frac{b \Lambda }{6} \notag \\ &+&\left(\frac{1}{u_R}+\frac{1}{u_S}\right)\frac{2 a \Lambda }{3}+\mathcal{O}({M^2,\Lambda^2,a^2, g_{h}^4},M a\Lambda,M a\Lambda  g_{h}^3) .
\end{eqnarray}
An  examination shows that this expression recovers many previous findings.   In the absence of the cosmological contributions, we get the results of the ordinary Hayward solutions investigated in \cite{JK}.   Moreover,   the  Schwarzschild black hole  and the  Kerr  black hole results   can be obtained by  sending  the extra parameters   to zero\cite{MAN}.
\section{Deflection angle  of light rays  by  cosmological Bardeen black holes }
 In this section,    we  deal with other  regular black holes. Precisely, we   study the deflection angle of light rays around the rotating Bardeen black holes with a  cosmological constant.    To start,   we give the associated metric   in  the   Boyer-Lindquist coordinates.  In this way, 
the  line element  of such a metric  reads as 
\begin{equation}
ds^{2}=-\frac{\Delta_{r}}{\Sigma}\left(dt-\frac{a \sin^{2}\theta}{\Xi}d\phi\right)^{2}+\frac{\Sigma}{\Delta_{r}}dr^{2}+\frac{\Sigma}{\Delta_{\theta}}d\theta^{2}+\frac{\Delta_{\theta} \sin^{2}\theta}{\Sigma}\left(a dt - \frac{r^{2}+a^{2}}{\Xi}d\phi\right)^{2}.
\end{equation}
This involves   the physical terms 
\begin{eqnarray}
\Delta_{r}&= &\left(r^{2}+a^{2}\right)\left(1-\frac{r^{2}\Lambda}{3}\right)-2M\left(\frac{r^{2}}{r^{2}+g_{b}^{2}}\right)^{\frac{3}{2}}r, \qquad  \Delta_{\theta}= 1+\frac{a^{2}\Lambda}{3}\cos^{2}\theta, \\  \notag
\Xi & = &1+\frac{a^{2}\Lambda}{3}, \qquad \Sigma = r^{2}+a^{2} \cos^{2}\theta
\end{eqnarray}
  where  $M$ and $a$ denote the mass and the rotating parameter of the black hole, respectively.  $g_b$  represents  the charge of the 
nonlinear electrodynamics. To compute the associated   deflection angle, the orbit equation  is needed  being relevant as indicated  in the previous section.   A close examination provides the following relation
\begin{eqnarray}
\left( \frac{du}{d\phi}\right)^{2} &=&\frac{1}{b^2}-\frac{4 a M u}{b^3}-u^2+2 M u^3+\frac{\Lambda }{3}-\frac{2 a \Lambda }{3 b^3 u^2}-\frac{8 a \Lambda  M}{3 b^3 u}-3 M u^5 g_b^2\notag\\ &+&\frac{6 a M u^3 g_b^2}{b^3}+\frac{4 a \Lambda  M u g_b^2}{b^3}  +    \mathcal{O}({M^2,\Lambda^2,a^2, g_{b}^3}).
\end{eqnarray}
After computations, we obtain 
\begin{eqnarray}
\phi_{RS} &=& \phi_{RS}^{Kerr}+\left(\frac{b^4 u_R^4+4 b^2 u_R^2-8}{2 b^3 \sqrt{1-b^2 u_R^2}}+\frac{b^4 u_S^4+4 b^2 u_S^2-8}{2 b^3 \sqrt{1-b^2 u_S^2}}\right) M g_{b}^2\notag \\ &-& \left( \frac{ 3 b^4 u_R^4-12 b^2 u_R^2+8}{ \left(1-b^2 u_R^2\right){}^{3/2}}+\frac{ 3 b^4 u_S^4-12 b^2 u_S^2+8}{\left(1-b^2 u_S^2\right){}^{3/2}}\right) \frac{ \Lambda  M g_{b}^2}{4}   +\left( \frac{6-3 b^2 u_R^2}{\sqrt{1-b^2 u_R^2}}+\frac{6-3 b^2 u_S^2}{\sqrt{1-b^2 u_S^2}}\right) \frac{a  M g_{b}^2}{b^4}\notag \\ &+& \left(\frac{ u_R}{ \sqrt{1-b^2 u_R^2}}+\frac{ u_S}{ \sqrt{1-b^2 u_S^2}}\right)\frac{b^3 \Lambda}{6} +\left(\frac{b \left(2-3 b^2 u_R^2\right)}{2 \left(1-b^2 u_R^2\right){}^{3/2}}+\frac{b \left(2-3 b^2 u_S^2\right)}{2 \left(1-b^2 u_S^2\right){}^{3/2}}\right)\frac{M\Lambda}{3}\notag \\ &+& \left( \frac{1-2 b^2 u_R^2}{u_R \sqrt{1-b^2 u_R^2}}+\frac{1-2 b^2 u_S^2}{u_S \sqrt{1-b^2 u_S^2}}\right) \frac{a\Lambda}{3}+ \mathcal{O}({M^2,\Lambda^2,a^2, g_{b}^3},M a\Lambda,M a\Lambda  g_{b}^2 ) .
\end{eqnarray}
To get the remaining terms appearing in the deflection angle  expression,   one should  find  the $\Psi$ terms. Concretely, we find
\begin{eqnarray}
\sin \Psi&=&b u-b M u^2+2 a M u^2 +\left( \frac{3b u^4 }{2} -3 a  u^4+\frac{b \Lambda   u^2}{4} \right)M g_b^2- \left(\frac{b M}{6} - \frac{a  }{3 u}+\frac{b }{6 u}\right)\Lambda \notag \\ &+&\mathcal{O}({M^2,\Lambda^2,a^2, g_{b}^3},M a\Lambda,M a\Lambda  g_{b}^2 ).
\end{eqnarray}
This relation  leads to
\begin{eqnarray}
\Psi_{R}&-&\Psi_{S}=\Psi_{R}^{Kerr}-\Psi_{S}^{Kerr}+\left(\frac{u_R^4}{\sqrt{1-b^2 u_R^2}}+\frac{u_S^4}{\sqrt{1-b^2 u_S^2}}\right)\frac{3 b M g_b^2}{2} \\ &-&\left( \frac{u_R^4}{\sqrt{1-b^2 u_R^2}}+\frac{u_S^4}{\sqrt{1-b^2 u_S^2}}\right) 3 a M g_b^2-\left(\frac{u_R^2 \left(2 b^2 u_R^2-1\right)}{\left(1-b^2 u_R^2\right){}^{3/2}}+\frac{u_S^2 \left(2 b^2 u_S^2-1\right)}{\left(1-b^2 u_S^2\right){}^{3/2}}\right) \frac{ b \Lambda  M g_b^2}{4} \notag \\ &-&\left( \frac{1}{u_R \sqrt{1-b^2 u_R^2}}+\frac{1}{u_S \sqrt{1-b^2 u_S^2}}\right) \frac{b \Lambda}{6}-\left( \frac{2 b^2 u_R^2-1}{\left(1-b^2 u_R^2\right){}^{3/2}}+\frac{2 b^2 u_S^2-1}{\left(1-b^2 u_S^2\right){}^{3/2}}\right) \frac{b \Lambda M }{6}\notag \\ &+&\left( \frac{1}{u_R \sqrt{1-b^2 u_R^2}}+\frac{1}{u_S \sqrt{1-b^2 u_S^2}}\right) \frac{a \Lambda}{3}+  \mathcal{O}({M^2,\Lambda^2,a^2, g_{b}^3},M a\Lambda,M a\Lambda  g_{b}^2 ).
\end{eqnarray}

Exploiting  the above equations, we obtain the  deflection angle of light rays
\begin{eqnarray}
\alpha_{b}&=& \left(\sqrt{1-b^2 u_R^2}+\sqrt{1-b^2 u_S^2} \right)\frac{2 M}{b}- \left( \sqrt{1-b^2 u_R^2}-\sqrt{1-b^2 u_S^2} \right)\frac{2 a M}{b^2}\notag \\ &-& \left( \frac{1-b^2 u_R^2}{u_R \sqrt{1-b^2 u_R^2}}+\frac{1-b^2 u_S^2}{u_S \sqrt{1-b^2 u_S^2}}\right) \frac{\text{b$\Lambda $}}{6}+\left( \frac{1}{\sqrt{1-b^2 u_R^2}}+\frac{1}{\sqrt{1-b^2 u_S^2}}\right)\frac{b \Lambda  M}{6}\notag \\ &+& \left(\frac{\sqrt{1-b^2 u_R^2}}{u_R}+\frac{\sqrt{1-b^2 u_S^2}}{u_S}\right) \frac{2 a \Lambda }{3}+\left(\frac{b^4 u_R^4+b^2 u_R^2-2}{\sqrt{1-b^2 u_R^2}}+\frac{b^4 u_S^4+b^2 u_S^2-2}{\sqrt{1-b^2 u_S^2}} \right) \frac{2 M g_b^2}{b^3}\notag \\ &-& \left(\frac{b^4 u_R^4+b^2 u_R^2-2}{\sqrt{1-b^2 u_R^2}}+\frac{b^4 u_S^4+b^2 u_S^2-2}{\sqrt{1-b^2 u_S^2}}\right) \frac{3 a M g_b^2}{b^4} +\left(\frac{5 b^2 u_R^2-8}{\sqrt{1-b^2 u_R^2}}+\frac{5 b^2 u_S^2-8}{\sqrt{1-b^2 u_S^2}}\right)\frac{\Lambda  M g_b^2}{4 b} \notag \\ &+&\mathcal{O}({M^2,\Lambda^2,a^2, g_{b}^3},M a\Lambda,M a\Lambda  g_{b}^2 )   .
\end{eqnarray}
This relation  can be simplified   by considering  acceptable  approximations.   Vanishing $u_Sb$ and $u_Rb$,    we can get an expression  involving the  divergent terms coupled to  the geometric    contributions  showing the AdS background dependence. The final  light deflection angle   expression is found to be  
\begin{eqnarray}
\alpha_{b} &=& \frac{4 M}{b}-\frac{4 a M}{b^2} -\frac{8 M g_b^2}{b^3}-\frac{4 \Lambda  M g_b^2}{b}+\frac{12 a M g_b^2}{b^4}+\frac{b \Lambda  M}{3}-\left(\frac{1}{u_R}+\frac{1}{u_S}\right)\frac{b \Lambda }{6}\notag \\&+&\left(\frac{1}{u_R}+\frac{1}{u_S}\right)\frac{2 a \Lambda }{3}+\mathcal{O}({M^2,\Lambda^2,a^2, g_{b}^2},M a\Lambda,M a\Lambda  g_{b}^2 ).  
\end{eqnarray}
As the above  black hole solution, we can recover the previous findings by  removing the extra  physical parameters.   The present expressions   can  generalize the previous results\cite{JK, MAN}. \\
Having obtained the deflection angle of light rays by regular rotating black holes in the presence of  the  cosmological constant,  we move to graphically  analysis the corresponding optical behaviors.  In particular, we  vary the involved parameters including the cosmological constant. Different regions of  the space parameter have been inspected and compared with the Kerr solutions. 
\section{Results and   discussions}
\begin{figure}[!ht]
		\begin{center}
		\centering
			\begin{tabbing}
			\centering
			\hspace{8.cm}\=\kill
			\includegraphics[width=8cm, height=7cm]{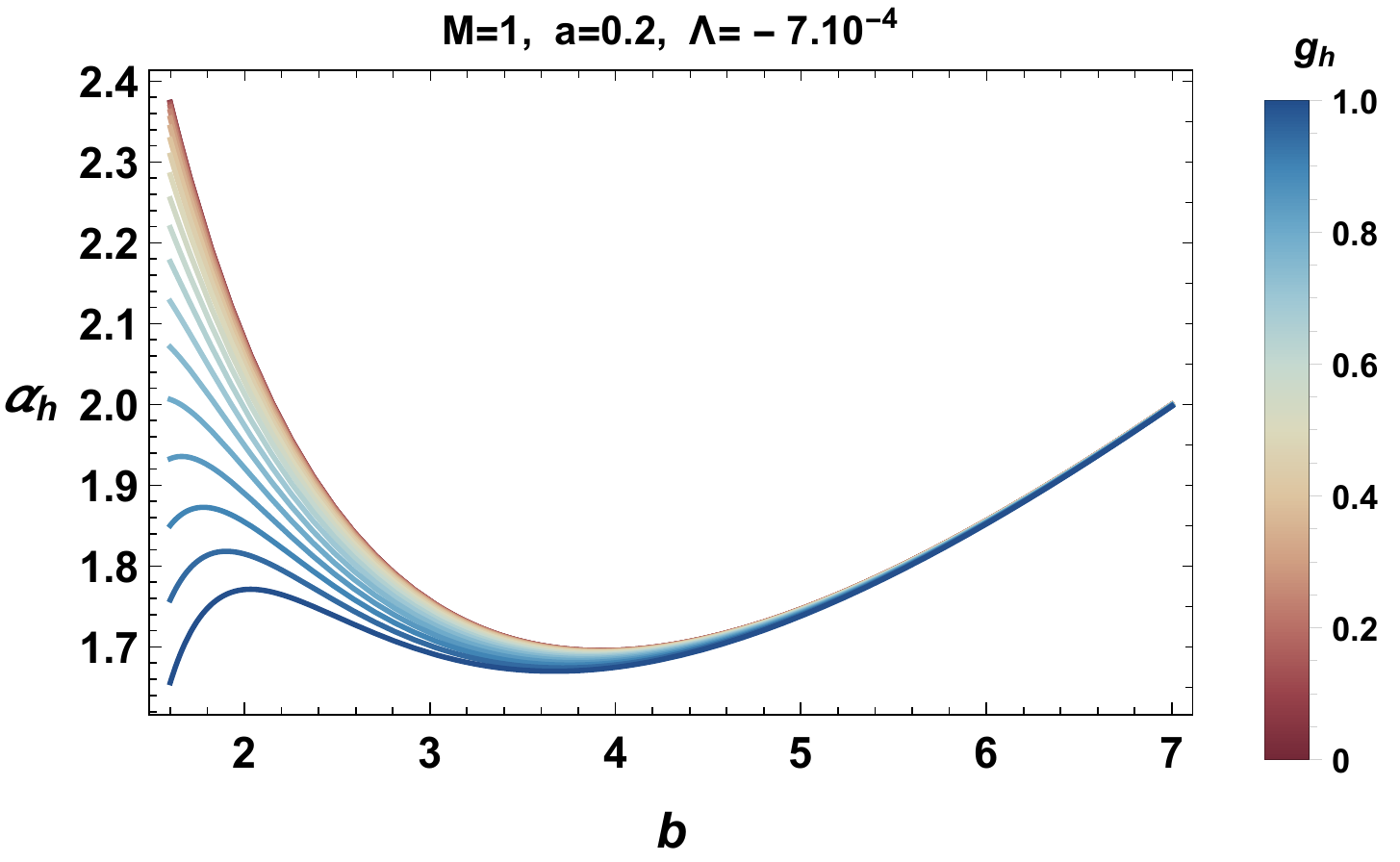} 
	\hspace{0.1cm}		\includegraphics[width=8cm, height=7cm]{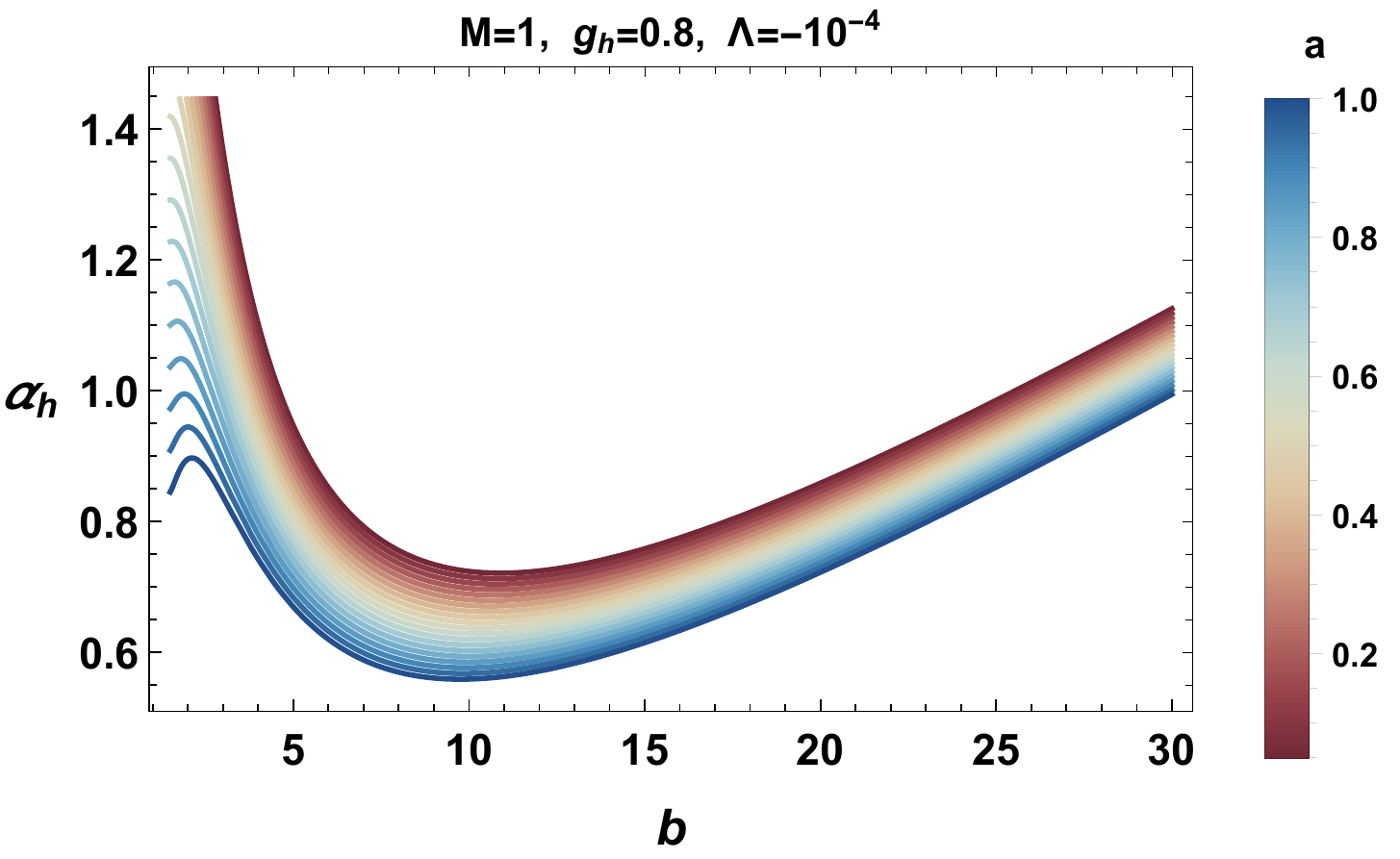}
		   \end{tabbing}
\caption{Deflection angle variation of Hayward  black holes with cosmological constant in terms of the impact parameter.}
\label{F1}
\end{center}
\end{figure}

To inspect  the  variation aspects of  the investigated  optical quantity,    the moduli space coordinated by the involved  parameters  is  needed. These behaviors  will be illustrated  by considering    certain regions of such a  parameter  space involving cosmological contributions.   Fixing  first  the black  hole  mass and  taking  a negative  cosmological  constant,   we will vary the   remaining ones.     For the  Hayward black hole  solutions, these aspects are presented in Fig.(\ref{F1}).

 For small values of  the impact parameter,  it has been remarked  that the deflection angle  of light rays  is a decreasing function with a remarkable   $g_h$   effect.  Indeed, it decreases with such a parameter.  For large values of the impact parameter, however,  the deflection angle of light rays  becomes  increasing  functions   with coinciding one dimensional real  curves for generic values of  $g_h$.   Moreover,  we observe similar  effect contributions appearing in the ordinary black hole solutions where the deflection angle  decreases  by increasing the rotation parameter.  The obtained results confirm  that the  space-time of the ordinary solutions   could be affected by $g_h$. 
\begin{figure}[!ht]
		\begin{center}
		\centering
			\begin{tabbing}
			\centering
			\hspace{8.cm}\=\kill
			\includegraphics[width=8cm, height=7cm]{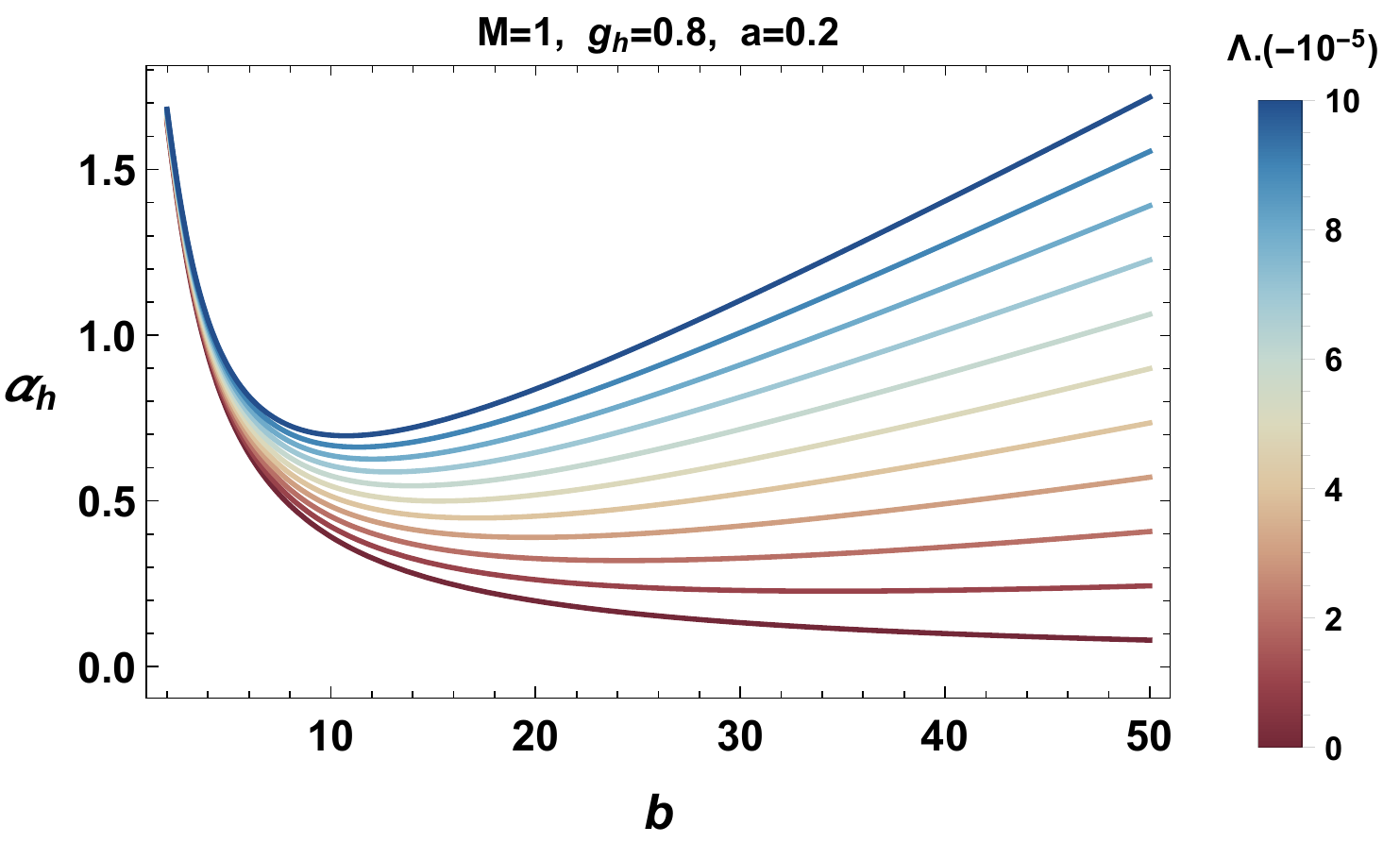} 
	\hspace{0.1cm}		\includegraphics[width=8cm, height=7cm]{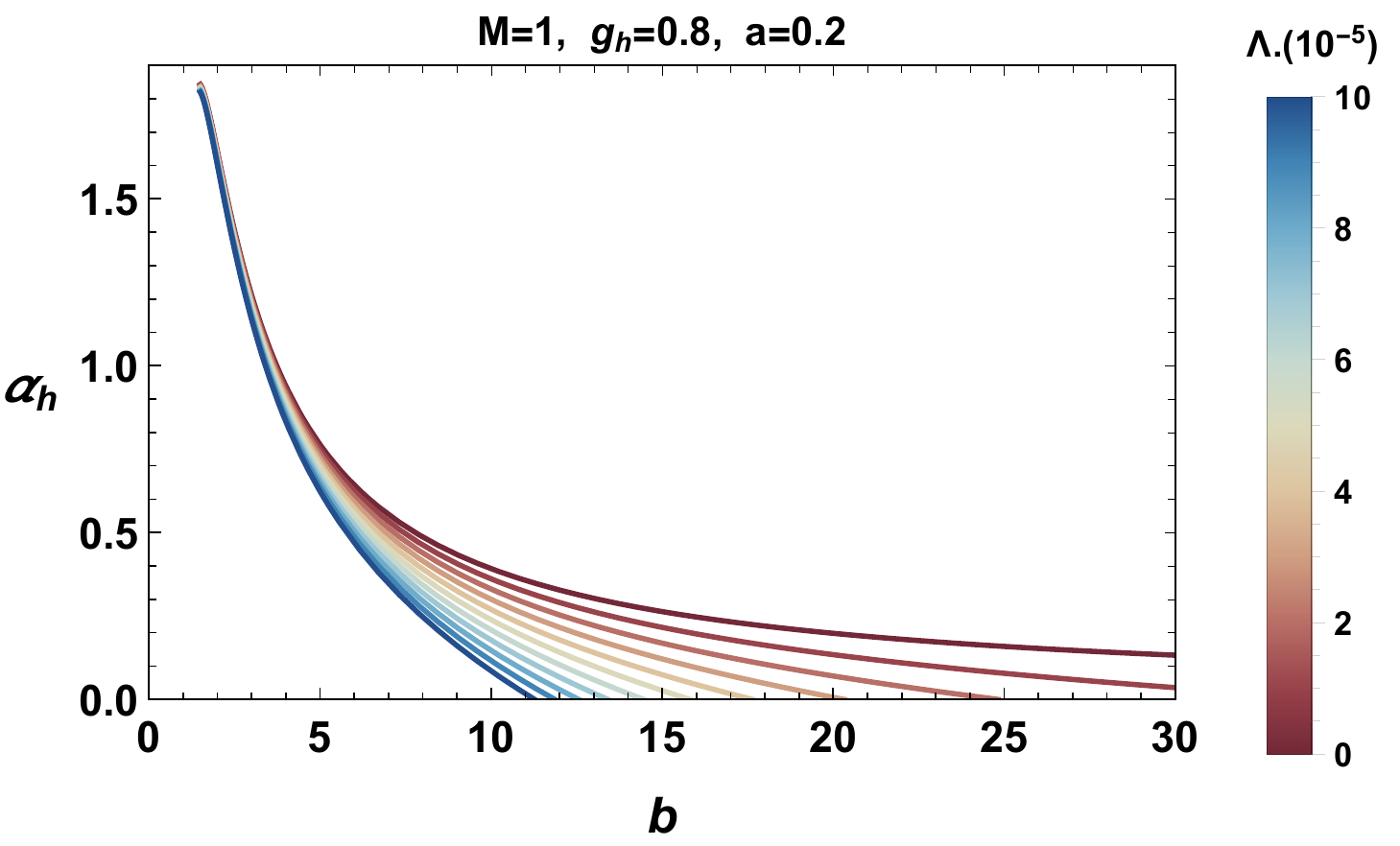}\\ 
		   \end{tabbing}
\caption{Deflection angle variations of Hayward  black holes in terms of the impact parameter $b$  for negative and positive values of  $\Lambda$.}
\label{F2}
\end{center}
\end{figure}
 To analyze the cosmological constant effect,   the  deflection  angle in terms of  $b$ for various values of $\Lambda$ will be inspected.  The   geometry pushes one to deal with different backgrounds  by considering negative and positive values of $\Lambda$. These behaviors are  presented  in Fig.(\ref{F2}).    The analysis depends on two regions associated with the  impact parameter values.   To do so, we consider first small values of the impact parameter. For small   negative  values of the cosmological constant,   we have remarked  a critical behavior where  the deflection angle  takes a minimal value which augments by decreasing  the cosmological constant $\Lambda$.  For positive values, however,  the deflection angle remains a decreasing  function of $b$.  For large values of the impact parameter,   we  have observed that the cosmological  constant effect is relevant.   Among others,  we remark a linear effect  of the deflection angle of light rays.

\begin{figure}[!ht]
		\begin{center}
		\centering
			\begin{tabbing}
			\centering
			\hspace{8.cm}\=\kill
			\includegraphics[width=8cm, height=7cm]{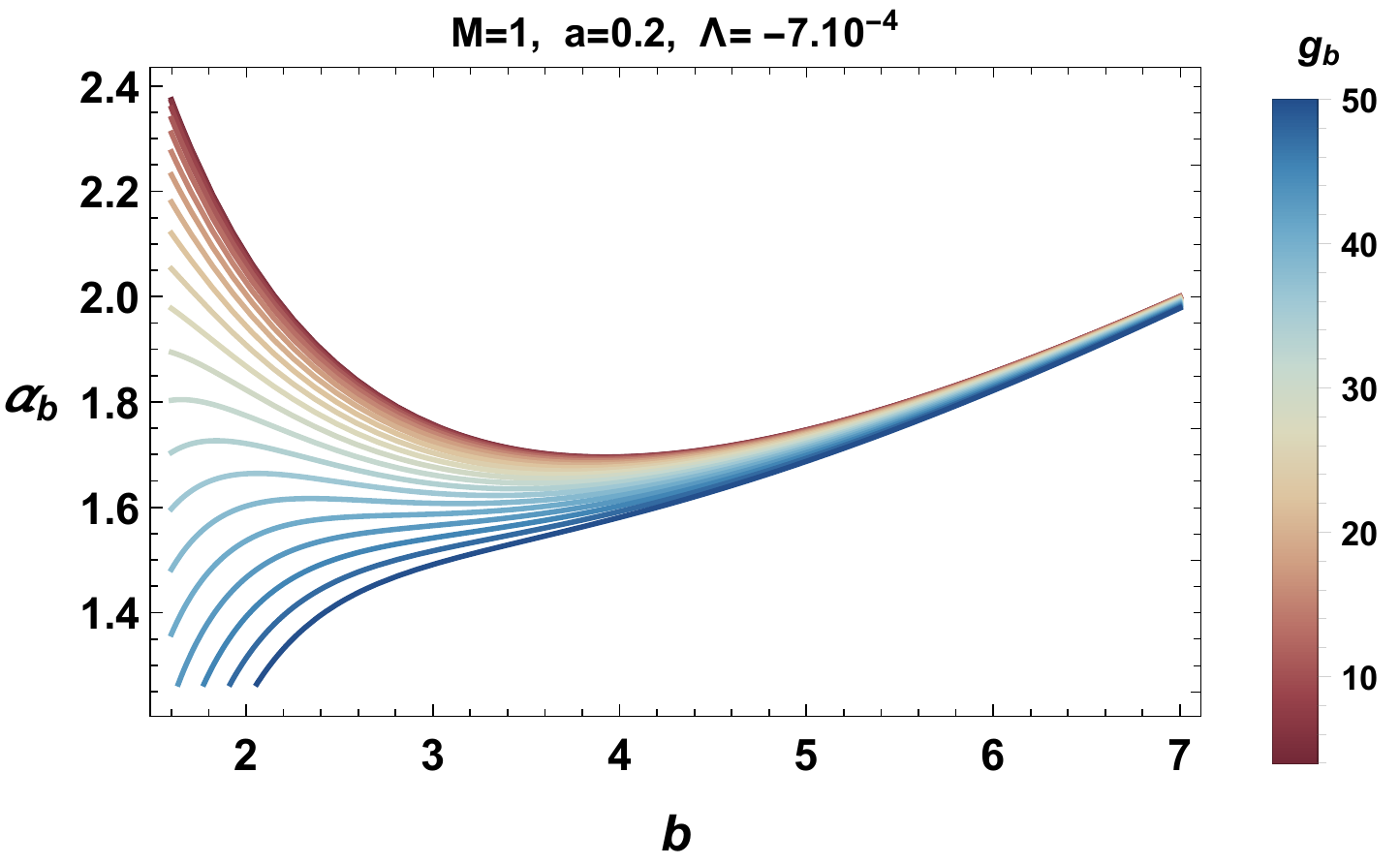} 
	\hspace{0.1cm}		\includegraphics[width=8cm, height=7cm]{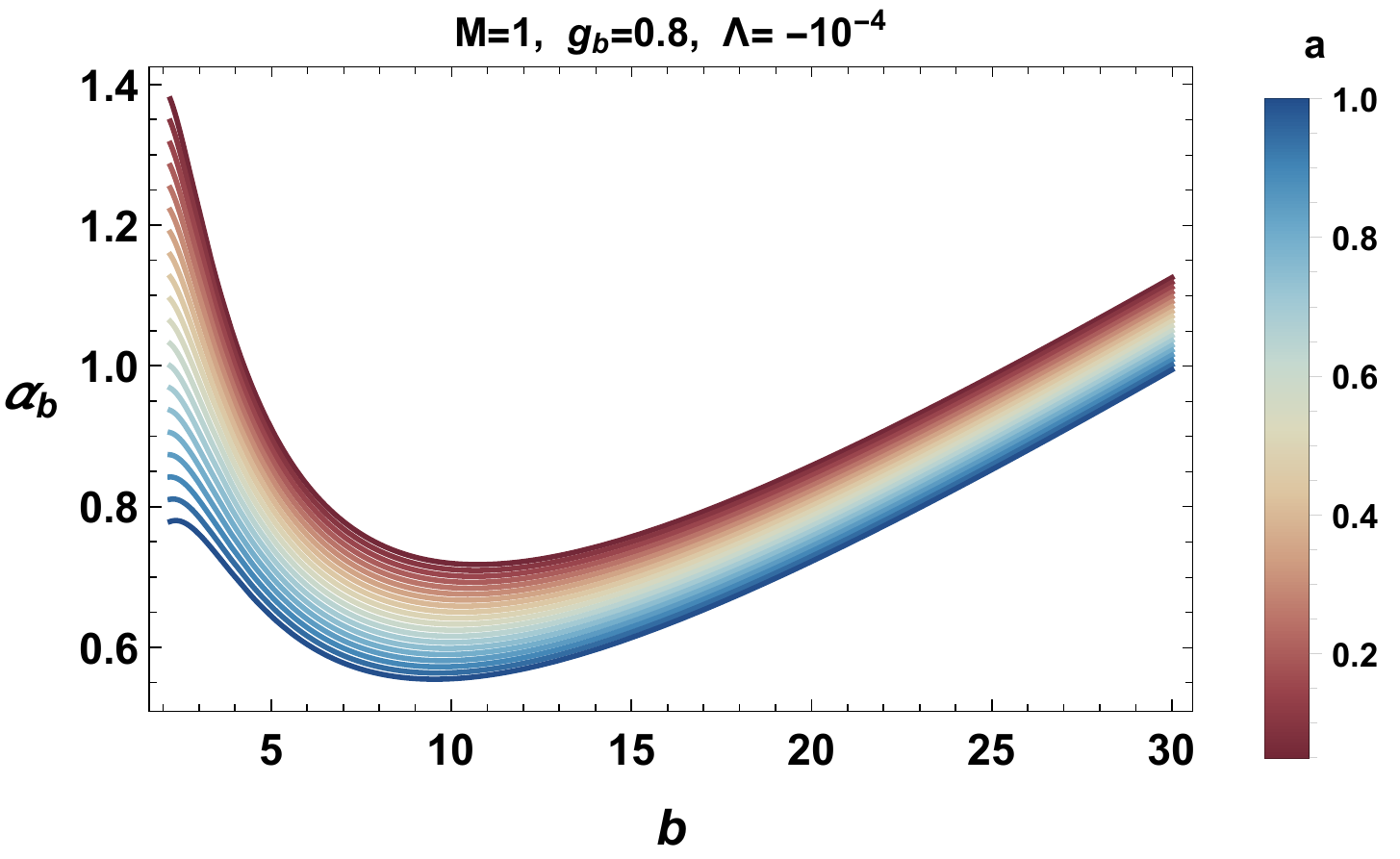}\\
	\includegraphics[width=8cm, height=7cm]{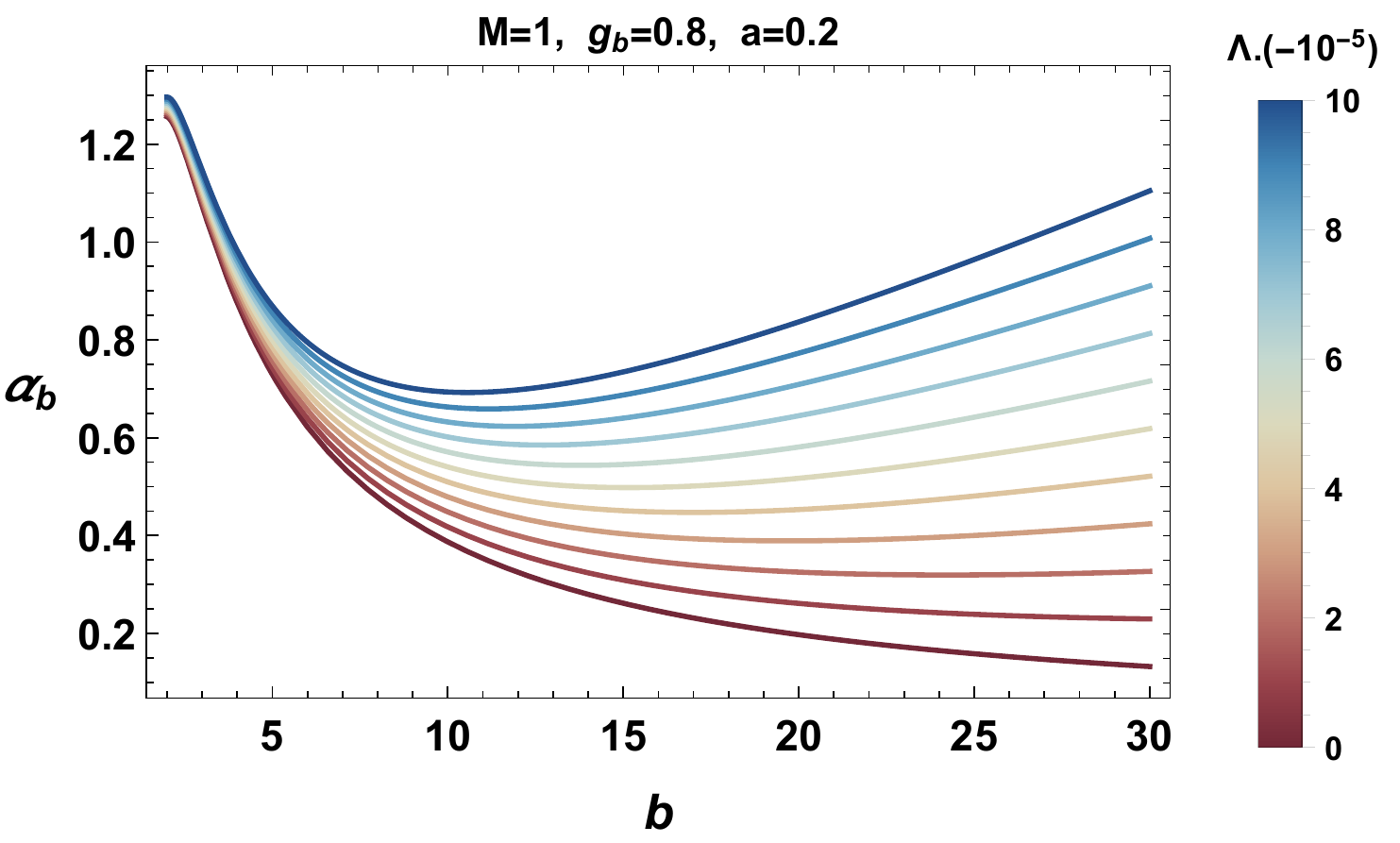} 
	\hspace{0.1cm}		\includegraphics[width=8cm, height=7cm]{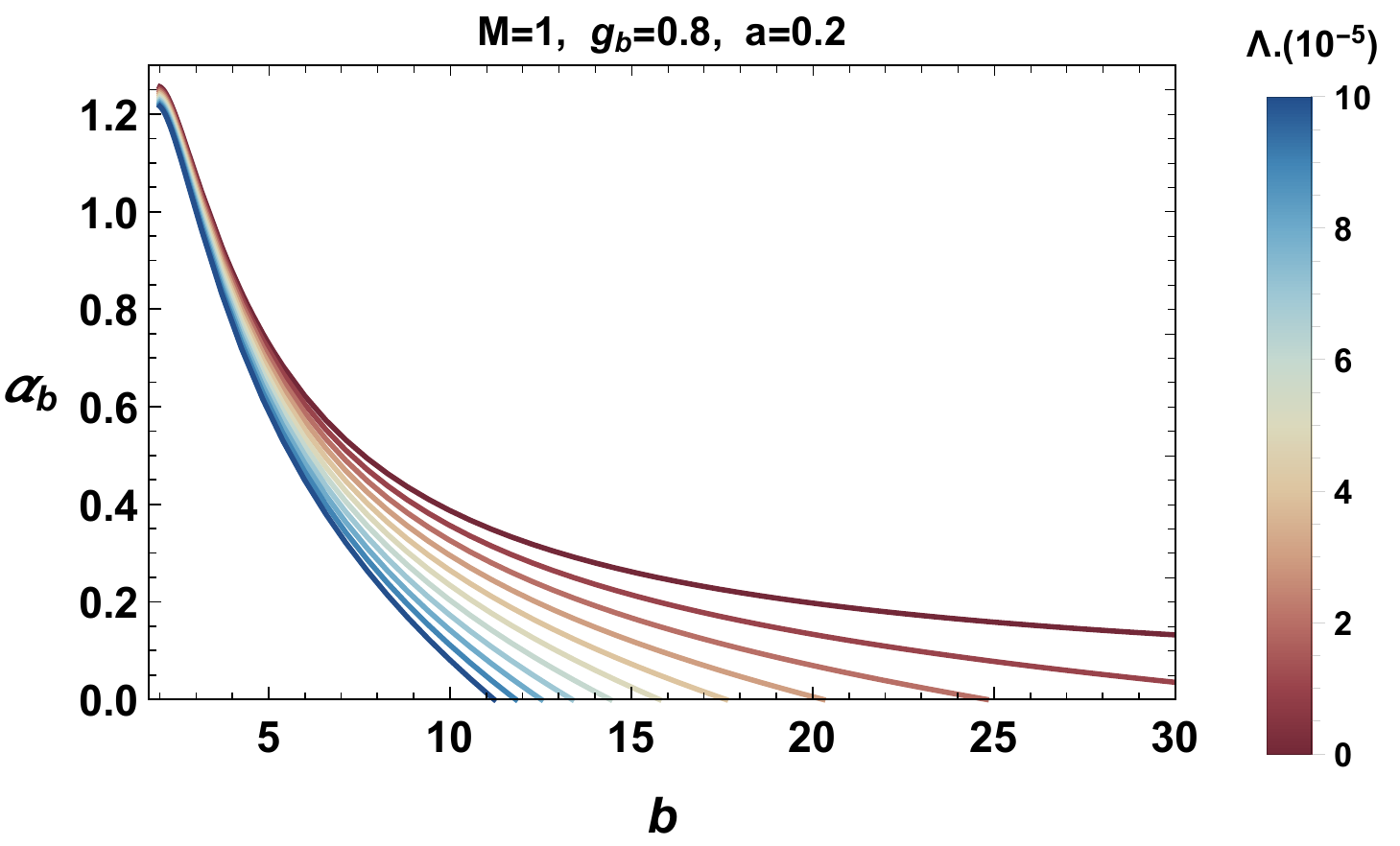}\\
		   \end{tabbing}
\caption{Deflection angle variations of Bardeen black holes with cosmological constant  in terms of the impact parameter.}
\label{F3}
\end{center}
\end{figure}

For the  Bardeen solutions, the  behaviors of the light deflection angle   are plotted in Fig.(\ref{F3}).

It has been observed   similar global  behaviors as the previous black hole solutions, where  we  have identified  the same parameter effects.  In particular,  the finding results  match perfectly with the fact that the  space-time of the ordinary solutions   could be affected by $g_h$.  However,  to unveil  certain  local distinctions between the  regular black hole backgrounds, we present a comparative  analysis. In  terms  of the Kerr solutions with cosmological contributions,  indeed,  the deflection angle of light rays can be reduced to  

\begin{eqnarray}
\alpha_{Kerr}^{\Lambda} \sim  \frac{4 M}{b}-\frac{4 a M}{b^2} +\frac{b \Lambda  M}{3}-\left(\frac{1}{u_R}+\frac{1}{u_S}\right)\frac{b \Lambda }{6}+\left(\frac{1}{u_R}+\frac{1}{u_S}\right)\frac{2 a \Lambda }{3}.
\end{eqnarray}

\begin{figure}[!ht]
		\begin{center}
			\includegraphics[width=8cm, height=7cm]{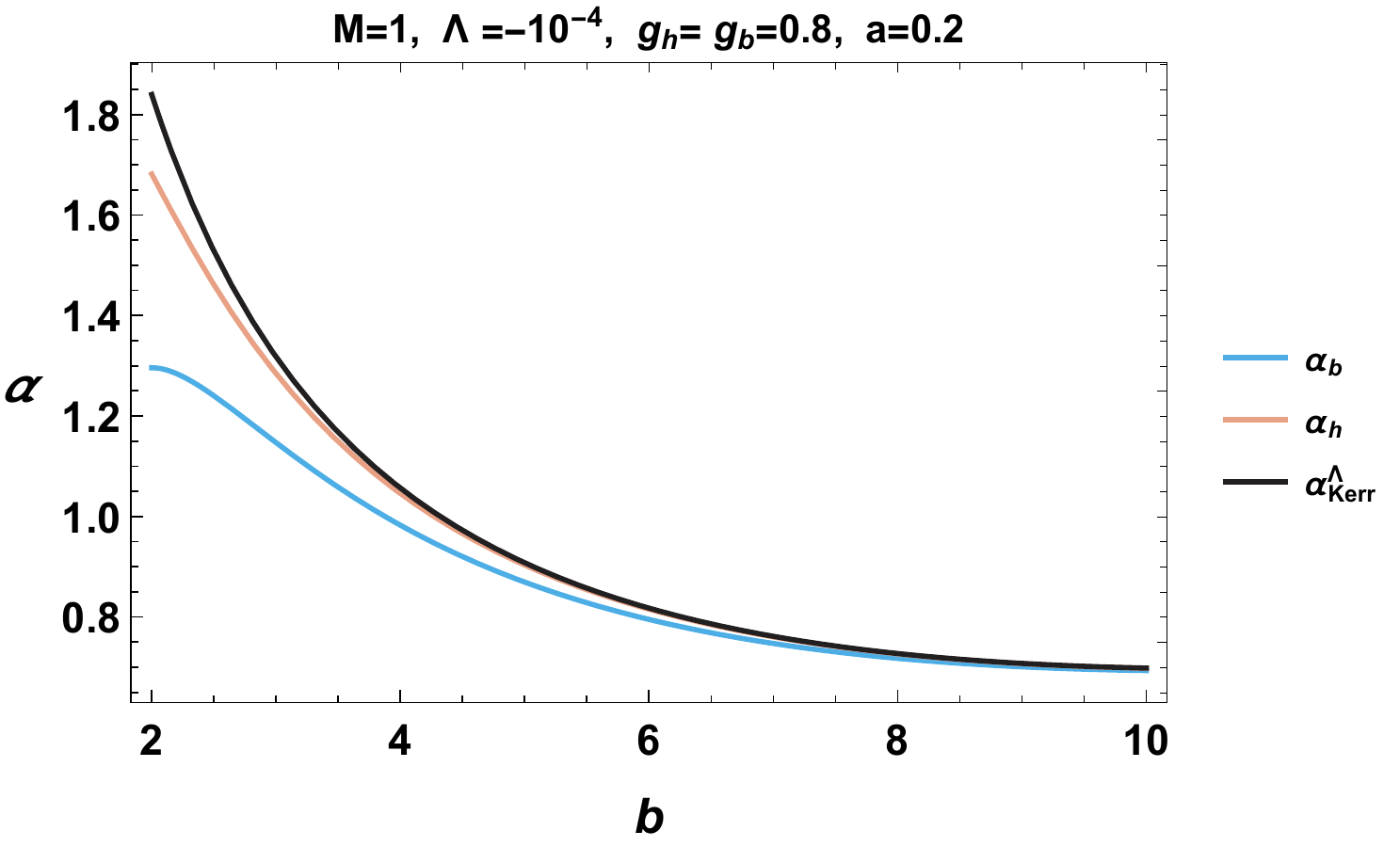} 
			\hspace{0.1cm}		\includegraphics[width=8cm, height=7cm]{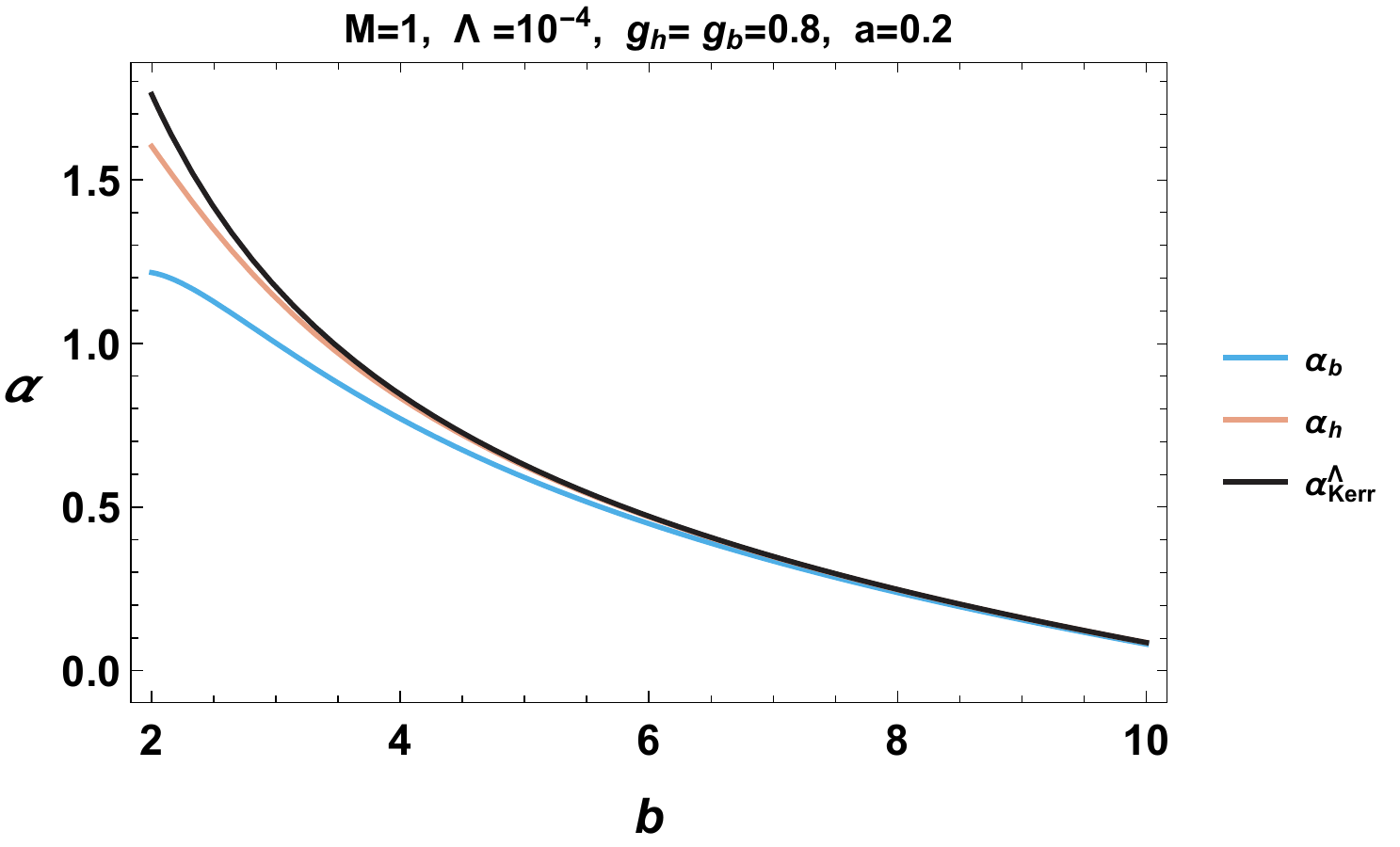}\\
\caption{Deflection angle variations of Bardeen, Hayward and Kerr black holes with a cosmological constant .}
\label{F4}
\end{center}
\end{figure}

  In Fig.(\ref{F4}),  we  illustrate  the behavior of the deflection angle of the light rays for  the regular black hole solutions   as a function of the impact parameter, compared with  the Kerr solutions.  
  It has been revealed  that small   ranges  of  $b$ are relevant in such an analysis for positive and negative cosmological constant values.   However,  large values do not bring notable distinctions due to coinciding real curves. It has been remarked that  the  negative  values of the cosmological constant  increase the light  deflection angle compared with positive ones. Indeed,  all solutions involve the same deflection angle of light rays. In this way, the discussion  concerns the first range.  Indeed, it has been remarked that the deflection angle of light rays by  the Bardeen  black holes is smaller than the one obtained for  Hayward solutions.  Moreover,  the light defection angle of both  solutions   is smaller than   the Kerr solution one. This result agrees with the previous one showing that   the  non-linear electrodynamic charges affect the space-time geometry.  Such parameters  decrease the deflection angle of light rays.
 We could  anticipate that  such an angle for both solutions goes to  the one of the the ordinary rotating  black holes with  the  cosmological constant.  In particular,  it  has been cheeked that  we can recover other  findings by sending the cosmological constant to zero\cite{JK,MAN,BW16}. 

 \section{Conclusion}
 
 In this    work, we have   studied the weak gravitational lensing  for  the  rotating regular black holes with cosmological contributions.    Considering  the photon motion in  the equatorial plane framework, we have   obtained the   optical metric  needed to  provide the associated   orbit equations.   Using   the Gauss-Bonnet theorem results,  we have   first  elaborated the  light  deflection angle     relation  of  the Hayward solutions by using   convenable  approximations.   The obtained expression  has recovered the previous findings.  Similar computations  have been applied to rotating  regular Bardeen black hole solutions.  The cosmological  relevant relations have been established  to derive the  deflection angle of light rays. As the first   regular model,  we have generalized certain previous works.  In particular,  it  has been remarked that  the non trivial electrodynamics charges affect the space-time geometry.  The obtained results have confirmed  the previous works.

     Moreover, some  other works have been recovered by  removing  such geometrical  contributions.
 To examine the deflection angle of light rays by such regular black holes with a cosmological constant, we have provided a graphical analysis by varying the involved relevant parameters. Fixing the mass parameter,   certain regions of the associated moduli space  have been  used by considering  negative and positive  values of the cosmological constant.  We have obtained extra corrections terms  being considered as extended contributions. Vanishing these terms, we have recovered  certain optical behaviors.  Finally, we have given  a comparative observation  concerning    regular black hole solutions with respect to the Kerr solutions.  It has been shown that large values of $b$ do not involve significant distinctions.  In the small ranges of $b$,   however,   the deflection angles  of the Bardeen  and  the Hayward  black holes are smaller than the  one of    the Kerr solutions with a cosmological constant. This finding matches perfectly  with the previous results indicating  that   the  non-linear electrodynamic charges affect the space-time geometry by decreasing  the deflection angle of light of rays.

This work  comes up with certain  open questions.  One  of them may concern  other methods being used to compute the deflection angle of light rays. It would be of interest to investigate such an optical data  by applying the  elliptic formalism exploited in many places.  We hope comeback to such an issue in future works.
\section*{Acknowledgments}
The authors would like to thank    N. Askour, Y. Hassouni,  K.  Masmar  and  M. B. Sedra for collaborations on related topics.  They would like
also to thank the editor and the  anonymous referee for  comments and
suggestions.
 This work is partially
supported by the ICTP through AF.

\end{document}